\documentclass{emulateapj}
\usepackage{graphicx}
\usepackage{epstopdf}
\usepackage{longtable}

\newcommand{\cii}{[C\,{\sc ii}]}
\newcommand{\ciii}{C\,{\sc iii}]}

\newcommand{\lya}{Ly$\alpha$}

\newcommand{\asec}{^{\prime\prime}}

\newcommand{\myemail}{chris.willott@nrc.ca}

\def\hst{{\it Hubble Space Telescope~}}

\def\spitzer{{\it Spitzer Space Telescope~}}

\def\co21{CO\,(2-1)}

\shorttitle{Star formation and the interstellar medium in $z>6$ LBGs}
\shortauthors{Willott et al.}

\begin{document}


\title{Star formation and the interstellar medium in $z>6$ UV-luminous Lyman-break galaxies}


\author{Chris J. Willott}
\affil{NRC Herzberg, 5071 West Saanich Rd, Victoria, BC V9E 2E7, Canada}
\email{\myemail}

\author{Chris L. Carilli}
\affil{National Radio Astronomy Observatory, P.O. Box 0, Socorro, NM 87801, USA}
\affil{Cavendish Astrophysics Group, University of Cambridge, Cambridge, CB3 0HE, UK}

\author{Jeff Wagg}
\affil{Square Kilometre Array Organization, Jodrell Bank Observatory, Lower Withington, Macclesfield, Cheshire SK11 9DL, UK}

\author{Ran Wang}
\affil{Kavli Institute for Astronomy and Astrophysics, Peking University, Beijing 100871, China}


%\altaffiltext{1}{Visiting Astronomer, Cerro Tololo Inter-American Observatory.
%CTIO is operated by AURA, Inc.\ under contract to the National Science


\begin{abstract}

  We present Atacama Large Millimeter Array (ALMA) detections of
  atomic carbon line and dust continuum emission in two UV-luminous
  galaxies at redshift 6. The far-infrared (FIR) luminosities of these
  galaxies are substantially lower than similar starbursts at later
  cosmic epochs, indicating an evolution in the dust properties with
  redshift, in agreement with the evolution seen in ultraviolet (UV)
  attenuation by dust. The \cii\ to FIR ratios are found to be higher
  than at low redshift showing that \cii\ should be readily detectable
  by ALMA within the reionization epoch. One of the two galaxies shows
  a complex merger nature with the less massive component dominating
  the UV emission and the more massive component dominating the FIR
  line and continuum. Using the interstellar atomic carbon line to
  derive the systemic redshifts we investigate the velocity of \lya\
  emission emerging from high-$z$ galaxies. In contrast to previous
  work, we find no evidence for decreasing \lya\ velocity shifts at
  high-redshift.  We observe an increase in velocity shifts from
  $z\approx 2$ to $z\approx 6$, consistent with the effects of
  increased IGM absorption.

\end{abstract}





\section{Introduction}

Observations from the \hst\ have given us a broad-brush picture of the
evolution of galaxies over cosmic time. Galaxies in the reionization
epoch at redshifts $z>6$ were rarer, smaller, bluer and had stronger
nebular emission lines than typical galaxies in the more evolved
universe \citep{Ono:2013,Bouwens:2014,Bouwens:2014a,Smit:2014}. These
observations can be explained by a Lambda Cold Dark Matter cosmology
with initial generations of stars forming out of low metallicity gas
in relatively low mass halos. However, many details are still to be
resolved.

Whether the early galaxy population is able to generate enough photons
to complete cosmic reionization by redshift 6 is still a matter of
debate and depends upon uncertainties such as the clumpiness of the
intergalactic medium (IGM) \citep{Finlator:2012}, the escape fraction
of ionizing photons from galaxies \citep{Siana:2015} and the UV
spectra of very low-mass galaxies \citep{Schaerer:2010}. Strong
evolution in the fraction of galaxies showing \lya\ emission
\citep{Pentericci:2011,Schenker:2012,Treu:2013} suggests a rapid
change in the IGM neutral hydrogen fraction at $z\sim 7$. Some of the
strength of this evolution could also be accounted for by increased
\lya\ absorption within galaxies due to lower \lya\ velocity shifts
\citep{Choudhury:2014}. The first measurements of \lya\ velocity shifts
in two $z>6$ galaxies do indeed appear to show smaller shifts than at
lower redshift \citep{Stark:2014a}.

Up til now most of our knowledge of high-redshift galaxies is based on
rest-frame ultraviolet (UV) and optical observations. Observations in
the far-infrared (FIR) can provide information on the physical
conditions of the interstellar medium (ISM) and dust and molecular gas
in obscured star-forming regions, as well as gas kinematics. At $z\sim
2$ FIR photometry of rest-frame UV-selected galaxies shows that the
star formation rate (SFR) is underestimated by a factor of 5 from UV
measurements alone \citep{Reddy:2012}. With the incredible sensitivity
of the Atacama Large Millimeter Array (ALMA) it should be possible to
detect dust continuum emission, molecular and atomic gas in
high-redshift galaxies to complete our understanding of their physical
properties \citep{Carilli:2013}. In addition, the line data provides
information on the gas kinematics and can be used to derive dynamical
masses, separate spatially unresolved mergers and define a systemic
redshift against which to measure the emergent \lya\ kinematics.

Several studies using ALMA and other facilities have resulted in
non-detections and claims that the FIR line and/or continuum fluxes
are lower than expected based on UV SFRs
\citep{Walter:2012,Kanekar:2013,Ouchi:2013,Ota:2014,Gonzalez-Lopez:2014,Maiolino:2015,Schaerer:2015}.
ALMA observations failed to detect either 1.2\,mm continuum or the
fine-structure line of singly-ionized carbon, \cii, in the $z>6.5$
galaxies {\it Himiko} and IOK-1 \citep{Ouchi:2013,Ota:2014}, despite
the strong UV continuum indicating SFR
$\sim 20 - 100 \,M_\odot$\,yr$^{-1}$. They showed that the FIR (cool
dust) contributions to the spectral energy distributions are similar
to nearby dwarf irregulars, more than an order of magnitude below the
levels expected for nearby starbursts/spirals such as M82 and M51. The
physical interpretation of the lack of dust and \cii\ emission in
these $z> 6$ galaxies is a very low metallicity ISM. Observations of
local low metallicity star-forming galaxies such as I Zw 18 also show
very low dust-to-gas and dust-to-stars ratios, further confirming such
galaxies as analogues of high-$z$ galaxies \citep{Fisher:2014}.

This picture has brightened recently with several
detections of the ISM in distant galaxies.  Although
\citet{Maiolino:2015} failed to detect three $6.8<z<7.1$ Lyman Break
Galaxies (LBGs) with ALMA in line or continuum, they did detect \cii\
emission at the expected redshift of BDF3299, offset by 4\,kpc from the UV
position, which they interpret as an accreting or satellite clump of
gas. \citet{Watson:2015} detected a gravitationally-lensed galaxy at
$z=7.5$ in dust continuum from which they infer $SFR_{\rm FIR}$ about
3 times greater than $SFR_{\rm UV}$, similar to the ratio typically
observed at $z\sim2$ \citep{Reddy:2012}. The lack of a \cii\ detection
 in this galaxy is likely due to an uncertain redshift, as
  Watson et al. note the lack
of an emission line redshift means the ALMA data only cover 50\% of
potential \cii\ line frequencies. \citet{Capak:2015} observed nine LBGs at
$5<z<6$ with ALMA and detected \cii\ emission from all of them and
dust continuum in four galaxies. They showed that the dust emission in
these galaxies is much weaker than expected based on similar galaxies
at lower redshift. The Capak LBGs were selected based on interstellar
UV lines so likely have higher metallicity than most
previously-targeted high-$z$ galaxies, although still lower than at
low redshift.

We are carrying out an ALMA program targeting $UV$-luminous LBGs at
$z>6$. In this paper we present observations of the first two galaxies
observed from our sample. The galaxies were first identified in
Canada-France-Hawaii Telescope (CFHT) optical and near-IR imaging and
both have spectroscopic redshifts based on a continuum break plus
\lya\ emission. The galaxy which we name here as CLM 1 was discovered
by \citet{Cuby:2003} at a redshift of $z=6.17$. The other galaxy, WMH
5 at $z=6.07$, was discovered in \citet{Willott:2013a}. Both galaxies
have near-IR continuum magnitudes $AB\sim 24$ making them among the
brightest rest-frame UV galaxies known at this epoch.
  
In Section 2 we describe our observations and in Section 3 our
results. Section 4 discusses the ratio of \cii\ to FIR continuum and
Section 5 the \lya\ velocity shifts observed. We draw conclusions in
Section 6.  Cosmological parameters of $H_0=67.8~ {\rm
  km~s^{-1}~Mpc^{-1}}$, $\Omega_{\mathrm M}=0.308$ and
$\Omega_\Lambda=0.692$ \citep{Planck-Collaboration:2015} are assumed
throughout.

\section{Observations}

\subsection{ALMA}

Observations were made during ALMA cycle 2, in June 2014, with between
29 and 32 antennas, and a maximum baseline of 650m. A total bandwidth
of 7.2 GHz was employed in Band 6 using 4 dual-polarization sub-bands
between 249 and 272 GHz. One of the sub-bands was centered at the
\cii\ line (rest frame 1900.5369 GHz) for each source. The rest of the
bands were used for continuum measurements. A total of 95 minutes
on-source integration time was obtained for each galaxy.

The initial data editing and calibration were performed as part of
standard data processing by the ALMA staff.  The calibrated visibility
data were then re-analyzed, performing additional flagging of bad time
periods and bad channels. The data were re-imaged using the CASA
Briggs weighting of the visibility data with Robust = 1 to create
continuum and channel images. For both sources, the Gaussian restoring
beam was close to circular, with a FWHM = 0.50". Spectral cubes at
15.625 MHz resolution ($17.5 \,{\rm km\,s}^{-1}$) were synthesized,
and smoothed spectrally for subsequent analysis, as required. The
continuum was subtracted in the image-plane, using the off-line
channels in the line cube.

The spectra were transformed from the observed topocentric system to
the local standard of rest (LSR). The velocity offsets applied were
$\approx -11 \,{\rm km\,s}^{-1}$.

Spectral and spatial Gaussian fitting was performed using the CASA
viewer, CASA fitting tools and custom software. Results for total line
fluxes, continuum flux densities, and parametric source sizes, are
given below. Flux uncertainties have 10\% added in
quadrature for absolute flux calibration uncertainty.

\subsection{Near- and Mid-Infrared}
 
To compare with the ALMA data we use near-infrared (NIR) data from the
ESO VISTA VIDEO survey \citep{Jarvis:2013}. Images from Data Release 3
in bands $Z, Y, J, H$ and $Ks$ were obtained, sampling the rest-frame
ultraviolet continua. Astrometric calibration of the VIDEO images to
the radio reference frame used by ALMA was performed by matching
bright VIDEO sources to the AllWISE catalog \citep{Cutri:2013} and
extracting the ALMA phase calibrator positions from the AllWISE
catalog. Given the size of the residuals this process gives an
uncertainty on astrometric frame matching of $\approx 0\farcs1$,
comparable to the positional uncertainties of the NIR data due to S/N
and seeing. Fluxes in the five NIR bands of the two target galaxies
were determined using aperture photometry with aperture corrections to
total fluxes. Due to the similar shapes, fluxes and S/N of the data in
the three bluest VIDEO bands, a deeper combined $ZYJ$ image of each
field was generated. Neither galaxy is significantly spatially
resolved in the NIR data.

\begin{figure*}
%\vspace{-0.5cm}
\includegraphics[angle=0,scale=0.60]{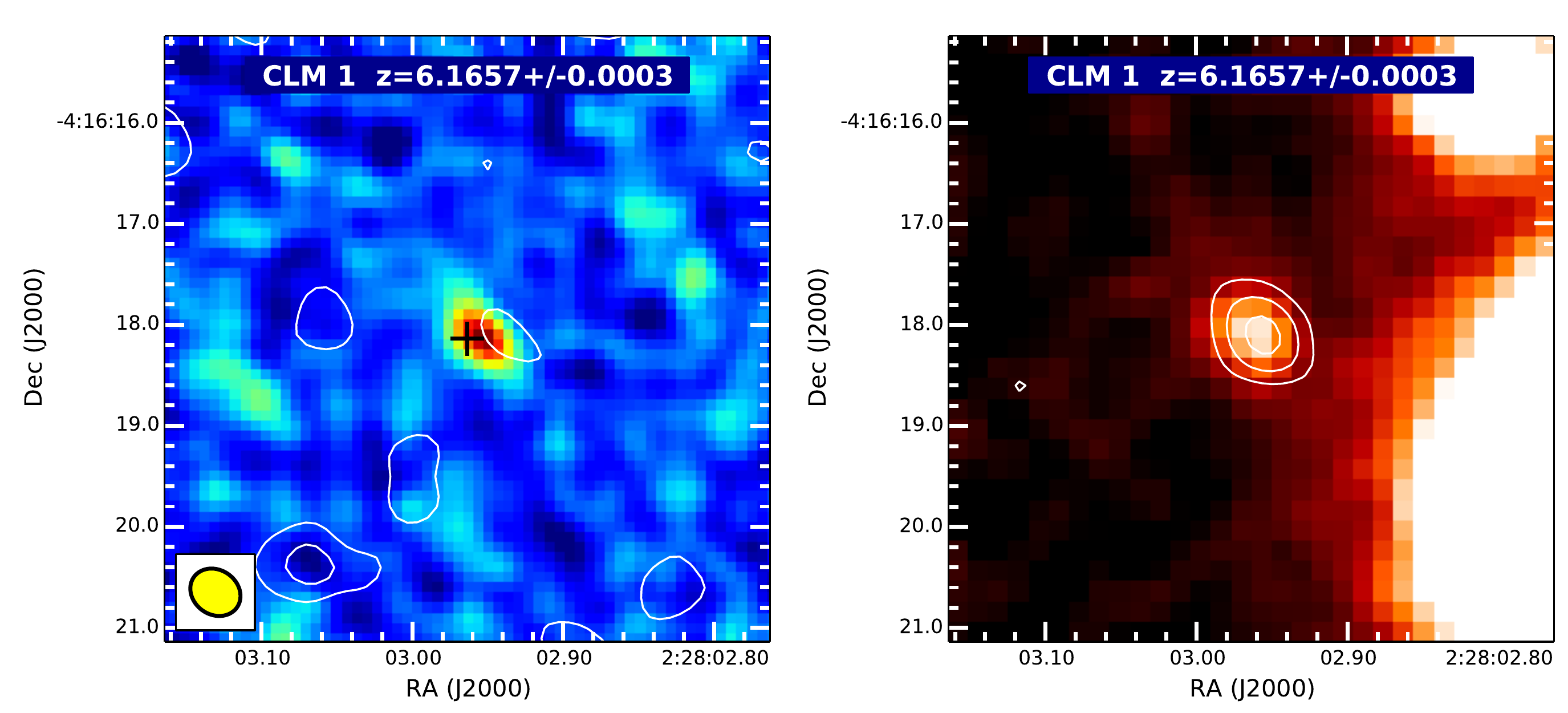}
\caption{{\it Left:} The background image is ALMA integrated \cii\
  line map of CLM 1. White contours are the 1.2\,mm continuum emission from the three line-free basebands at contour levels 1.5,\,2.5 $\sigma$\,beam$^{-1}$, specially chosen to show the low S/N possible continuum detection close to the \cii\ emission. The near-infrared centroid is plotted as a black plus symbol. The restoring beam is shown in yellow.\\ {\it Right:} The background is the $zYJ$ NIR image. Contours show the ALMA \cii\ emission from the left panel at levels 3,\,5,\,7\,$\sigma$\,beam$^{-1}$. The rest-frame UV continuum and \cii\ emission are co-spatial. Two foreground galaxies are visible on the right side of the image.} 
\label{fig:mapsclm1}
\end{figure*}

\begin{figure}
\hspace{-0.1cm}
\includegraphics[angle=0,scale=0.32]{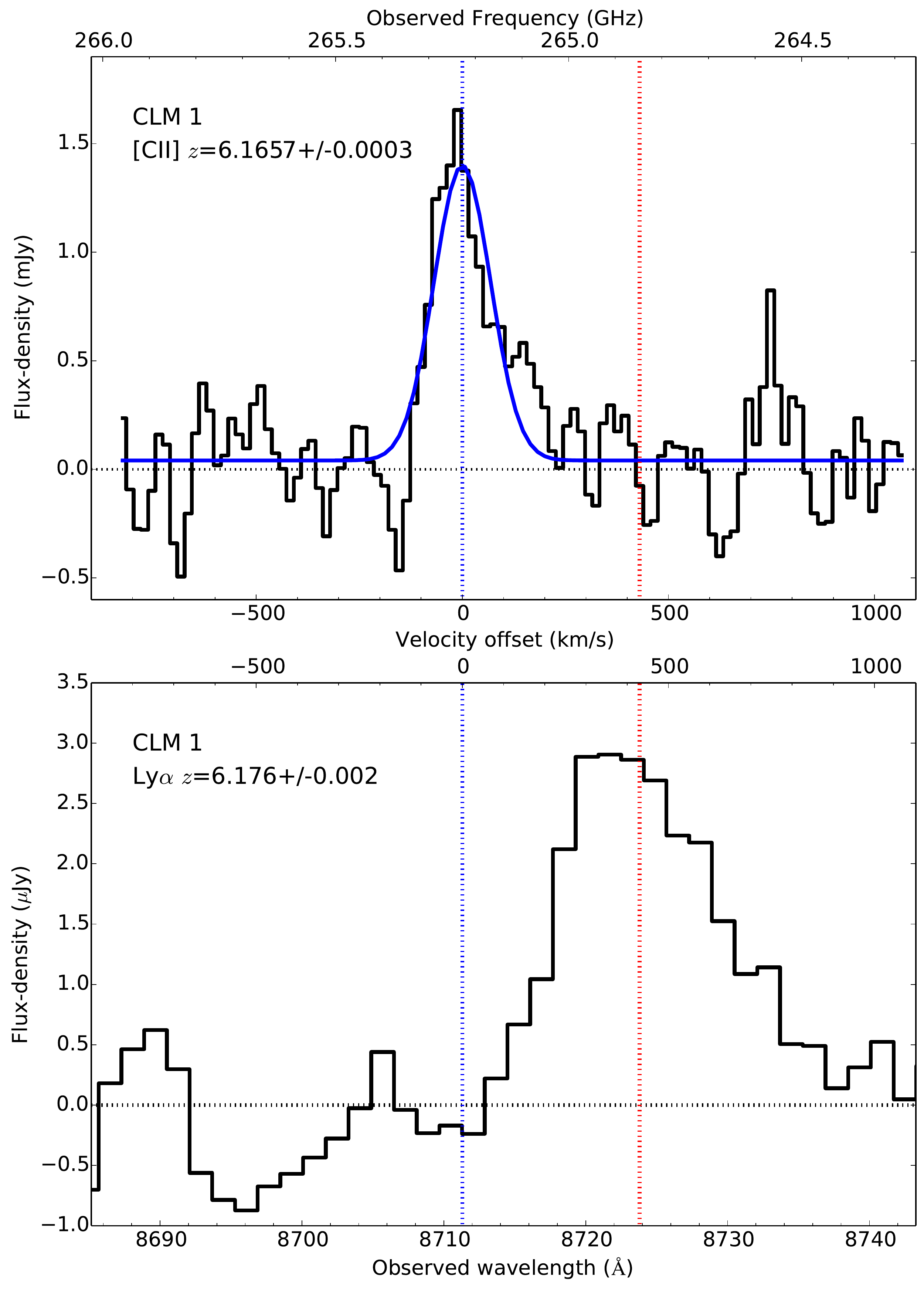}
\caption{ALMA \cii\ (upper) and VLT FORS2 \lya\ (lower) spectra of CLM
  1. The ALMA spectrum is well fit by a single Gaussian emission line, with a marginal red wing excess. The \lya\ spectrum shows typical blue-absorbed asymmetry. Vertical blue and red dotted lines mark the \cii\ Gaussian fit peak and \lya\ flux-weighted centroid, respectively. \lya\ is offset from \cii\ by $+430 \pm 69$ km\,s$^{-1}$.}
\label{fig:linespecclm1}
\end{figure}

The two galaxies have been observed with the \spitzer in the IRAC 3.6
and 4.5$\mu$m bands during the SERVS survey \citep{Mauduit:2012}. Due
to the proximity of bright galaxies along the line-of-sight to both
$z>6$ galaxies it was necessary to perform simultaneous fitting of all
objects in the IRAC field using the $ZYJ$ VIDEO image as a prior for
the object shapes. Models were fit to the NIR images using GALFIT
\citep{Peng:2010} and subsequent fitting of the IRAC data was done
with PyGFIT \citep{Mancone:2013}. Both $z>6$ galaxies are clearly
detected at both 3.6 and 4.5$\mu$m, although with flux uncertainties
of $\sim 30$\%.

\subsection{Lyman-$\alpha$ spectroscopy}

Optical spectroscopy, including the detection of asymmetric \lya\
emission lines and continuum breaks, of both galaxies were presented
in their discovery papers \citep{Cuby:2003,Willott:2013a}. In order to
compare these spectra to the ALMA spectra they were corrected from
observed wavelength to the LSR by applying velocity corrections of
$+26 \,{\rm km\,s}^{-1}$ and $+32 \,{\rm km\,s}^{-1}$ for CLM 1 and
WMH 5, respectively. The spectrum of CLM 1 was obtained using FORS2 at
the ESO VLT with a spectral resolution of $R=1400$ and that of WMH 5
using GMOS at Gemini-North with $R=1000$.
\section{Results}

\subsection{CLM 1}

For both galaxies there are strong detections of the \cii\ line
evident in the datacubes at close to the expected position and
velocity. In Figure \ref{fig:mapsclm1} we present the ALMA and NIR
imaging of the galaxy CLM 1. The \cii\ map is made from a sum over all
channels showing significant line emission. The \cii\ emission is
spatially extended along a similar direction as the beam extension. A
Gaussian fit to the source gives an observed size of $0\farcs73 \times
0\farcs42$ at position angle east of north (PA) = 42, compared with a
beam size of $0\farcs52 \times 0\farcs44$ at PA = 55. The CASA fitting
software does not provide a deconvolved source size due to the only
slightly larger size than the beam and outputs that the intrinsic
source may be as large as $0\farcs55 \times 0\farcs09$. 

The white contours in Figure \ref{fig:mapsclm1} show the 1.2 mm
continuum emission as measured from the 3 sub-bands without the \cii\
line. In contrast to the strong and clear \cii\ emission the continuum
in this source is very weak. There is a marginal continuum detection
at the $2\sigma$ level that is only considered as plausible true
emission because its centroid co-incides to within $0\farcs3$ of the
\cii\ line centroid. There are several other peaks of this magnitude
within the $6\asec \times 6\asec$ image of Figure \ref{fig:mapsclm1}
that may be due solely to noise. We note that there is an
emerging trend in ALMA observations of non-ULIRG high-redshift
galaxies that the \cii\ line emission is detected at much higher
significance than the continuum, despite the large instantaneous
bandwidth of 7.5 GHz \citep{Riechers:2014,Willott:2013,Willott:2015,Capak:2015}. We
discuss the ratio of line to continuum emission further in Section
\ref{ciifir}.

The NIR image in Figure \ref{fig:mapsclm1} shows the location of the
rest-frame UV continuum in CLM 1. The \lya\ contribution to this flux
is negligible. The galaxy is indistinguishable from a point source at
the $0\farcs8$ resolution of the NIR data. The centroids of the \cii\
and rest-frame UV co-incide to within $0\farcs1$, suggesting they
trace the same star forming regions of the galaxy, at this resolution.

Visible to the east of CLM 1 is a lower redshift elliptical galaxy
whose centroid is only $3\asec$ from CLM 1. The presence of this
galaxy combined with the high UV luminosity of CLM 1 raises the
possibility that the high luminosity is in part due to magnification
by gravitational lensing. To determine the potential lensing
magnification we have analyzed the properties of this elliptical
galaxy. We fit galaxy models to 12 band photometry from the CFHT
Legacy Survey, VIDEO and AllWISE using the FAST code
\citep{Kriek:2009}.  The spectral energy distribution (SED) is
well fit by an old stellar population at $z=0.54$ with a stellar mass
of $5\times 10^{10}~(M_\odot)$. Using the observed correlation between
velocity dispersion and stellar mass \citep{Wake:2012} at $z=0.1$ and
with negligible evolution from $z=0.5$ to $z=0.1$ \citep{Shu:2012}
this gives a velocity dispersion of 126 km\,s$^{-1}$. Assuming an
isothermal sphere potential this lensing configuration provides a
magnification factor of only 1.13. Given this relatively low
magnification we do not make any lensing corrections to physical
values in this paper.

Figure \ref{fig:linespecclm1} plots the \cii\ spectrum of CLM 1 and
the \lya\ spectrum for comparison. Interpretation of the velocity
difference between \cii\ and \lya\ is deferred to Section
\ref{lyashifts}. The \cii\ spectrum is well fit by a single Gaussian
plus a very low level, flat continuum. The fit continuum level is
consistent with the low S/N continuum detection from the other 3
sub-bands described above. There is marginal excess emission in the
red wing. From the Gaussian peak frequency we determine the systemic
redshift of the galaxy as $z=6.1657 \pm 0.0003$. The Gaussian FWHM is
$162 \pm 23$ km\,s$^{-1} $. Table 1 contains further measurements made
from the ALMA data.

With this velocity and the upper limit on the source size
($<0\farcs55$ or 3.2\,kpc) we can calculate an upper limit to the
dynamical mass.  Following the procedure of \citet{Wang:2013} we
derive $M_{\rm dyn}< 6.9\times 10^9 / \sin^2i \,M_\odot$ where $i$ is the
unknown inclination angle.

Using photometry from the rest-frame UV, optical and far-IR we fit
galaxy spectral models to understand the evolutionary state of CLM
1. A synthetic ALMA continuum {\it filter} is generated using the
  wavelengths covered by the three line-free sub-bands. 
We use the Python implementation of the CIGALE package
\citep{Roehlly:2012} to fit the SED. This code determines the
ultraviolet-optical attenuation and its re-emission in the
infrared. Stellar light is modeled with a single \citet{Bruzual:2003}
stellar population with constant star formation rate, Chabrier initial
mass function and metallicity = 0.008. Dust attenuation followed the
\citet{Calzetti:2000} dependence on wavelength with re-emission in the
IR via the SED parameterization of \citet{Dale:2014}. 

\begin{figure}
\hspace{-0.35cm}
\includegraphics[angle=0,scale=0.48]{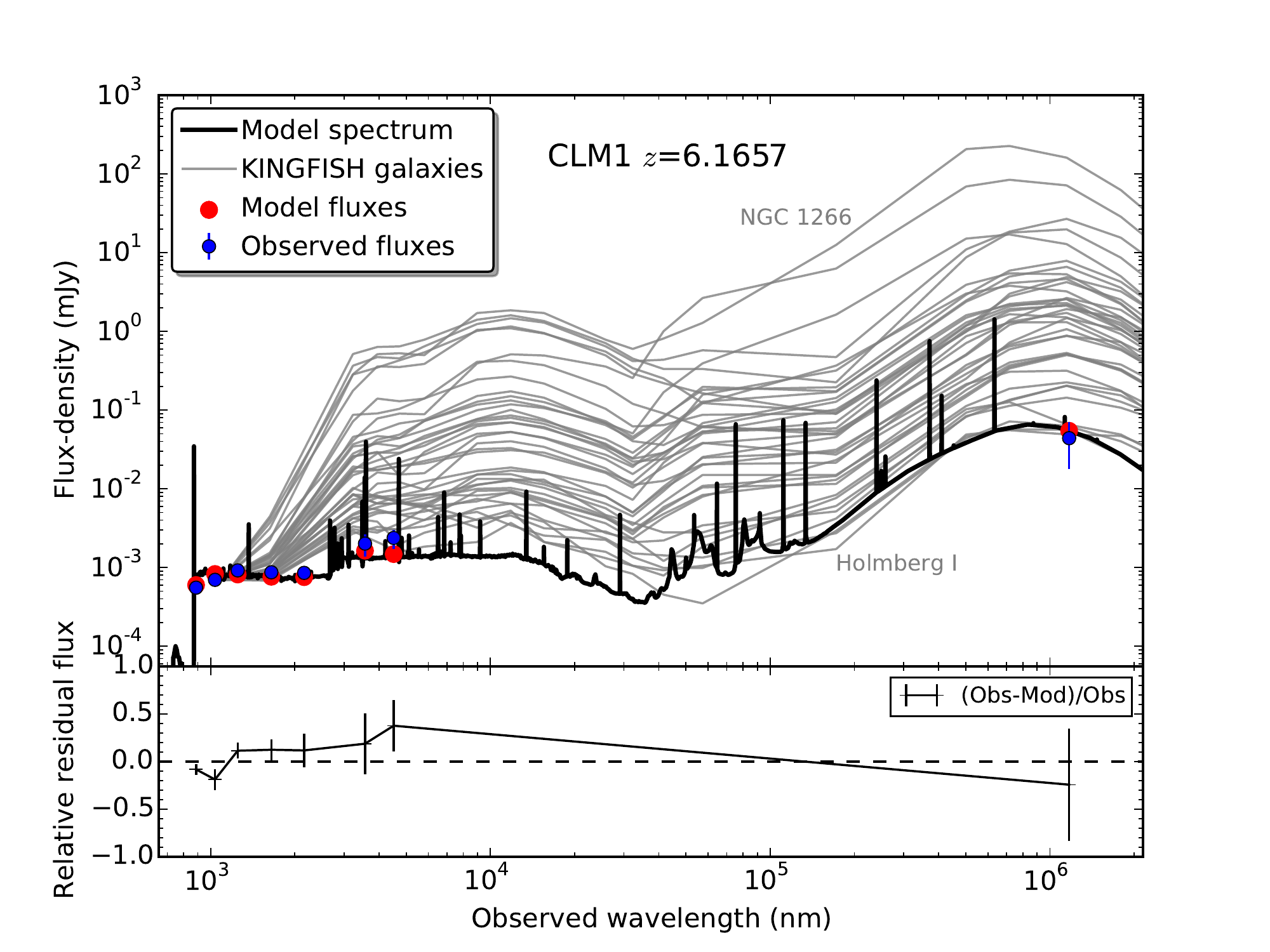}
\caption{Observed-frame optical to far-IR SED of CLM 1 (blue circles). The best-fit model from {\small CIGALE} is shown as a black curve and the model fluxes through the observed filters as red circles. The grey curves are 39 nearby galaxy SEDs with UV to far-IR photometry from the KINGFISH survey \citep{Dale:2012} ranging from dusty infra-red galaxies to dwarf irregulars with very low thermal dust emission. The lower panel shows the residuals from the best-fit model. CLM 1 has a SED most similar to the least dusty nearby dwarf irregulars.}
\label{fig:sedclm1}
\end{figure}

\begin{table}
\begin{center}
\caption{Millimeter data for the $z>6$ LBGs\label{tab:data}}
\vspace{-0.5cm}
\begin{tabular}{lll}
\tableline
& CLM 1 & WMH 5 \\
\tableline
%\lya\ redshift 
$z_{\rm Ly\alpha}$ &  $6.176 \pm 0.002 $ &  $6.076 \pm 0.001 $ \\
%\cii\ redshift 
$z_{\rm [CII]}$ &  $6.1657 \pm 0.0003$ & $6.0695 \pm 0.0003\,^{\rm a}$\\
                    &                                      & $6.0639 \pm 0.0003\,^{\rm b}$\\
FWHM$_{\rm [CII]}$ & $162 \pm 23$ km\,s$^{-1} $& $251 \pm 15$ km\,s$^{-1}\,^{\rm a}$\\
                            &                                              & $68 \pm 13$ km\,s$^{-1}\,^{\rm b}$\\
%\cii\ line flux 
$I _{\rm [CII]} ~($Jy\,km\,s$^{-1}) $ & $0.234  \pm 0.031$ & $0.659 \pm 0.071$\\
%\cii\  luminosity 
$L_{\rm [CII]} ~(L_\odot)$ & $(2.40 \pm 0.32) \times 10^8$  & $(6.60  \pm 0.72) \times 10^8$\\
%mm flux-density 
$f_{\rm 1.2mm}\ (\mu$Jy) & $44 \pm 26$ & $218 \pm 41$\\
%Far-IR luminosity 
$L_{\rm FIR} ~ (L_\odot)$ & $(2.5 \pm 1.5) \times 10^{10}$  & $(1.25 \pm 0.24) \times 10^{11} $\\
$L_{\rm [CII]} / L_{\rm FIR}$ & $(9.5 \pm 5.7) \times 10^{-3}$  &  $(5.3 \pm 1.2) \times 10^{-3}$  \\ 
SFR$_{\rm [CII]}\,(M_\odot\,{\rm yr}^{-1})$  & $24 \pm 3$  & $66 \pm 7$\\
SFR$_{\rm FIR}\,(M_\odot\,{\rm yr}^{-1})$  & $7 \pm 4$  & $33 \pm 6$\\
SFR$_{\rm SED}\,(M_\odot\,{\rm yr}^{-1})$  & $37 \pm 4$  & $43 \pm 5$\\
\tableline
\end{tabular}
\end{center}
{\sc Notes.}---\\
$^{\rm a}$ The redshift and FWHM of component `A' only.\\
$^{\rm b}$ The redshift and FWHM of component `B' only.\\
Uncertainties in $L_{\rm FIR}$ and SFR only include measurement uncertainties, not the uncertainties in extrapolating from a monochromatic to integrated luminosity, luminosity-SFR calibrations, or uncertainty in star formation history.
\end{table}

The highest likelihood parameters are a stellar mass of $1.3 \times
10^{10} \,M_\odot$, SFR\,$=37 \,M_\odot {\rm yr}^{-1}$ and minimal
dust attenuation of E(B-V)=0.01. This very low amount of dust
attenuation is constrained by the blue rest-frame UV slope of
$\beta=-2.0$ (where$f_\lambda \propto \lambda^{\beta}$) and in an IR
remission scenario consistent with the low FIR luminosity from the
marginal 1.2\,mm continuum detection. We note that an increasing star
formation rate and/or higher fraction of the IRAC flux contributed by
nebular emission lines could substantially decrease the stellar mass,
bringing it more into line with the dynamical mass upper limit. It is
also possible that the \cii-emitting gas is not tracing the full
dynamical mass of the system.

In Figure \ref{fig:sedclm1} we show the observed photometry of CLM 1
along with the best-fit model photometry, the model spectrum and, in
the lower panel, the residuals from the fit. With almost as many free
parameters as data points, a good fit has been found. The observed IRAC
fluxes are slightly higher than the model suggesting possibly stronger
nebular lines than modelled, although there are large uncertainties in
the IRAC fluxes. The gray curves in Figure \ref{fig:sedclm1} show
redshifted SEDs for nearby galaxies from the KINGFISH survey,
normalized at rest-frame 150\,nm. These galaxies have been observed
with {\it GALEX, Spitzer Space Telescope, Herschel Space Observatory}
and in the optical and near-IR \citep{Dale:2007,Dale:2012}. The
galaxies range in total IR luminosity from $10^7$ to
$10^{11}\,L_\odot$. CLM 1 has a SED most similar to the $\sim 10^7
\,L_\odot $ galaxies, despite having an IR luminosity of $> 10^{10}
\,L_\odot $, in essence it is a scaled up (by $10^3$) version of the
low metallicity dwarf irregulars.

It has recently been shown that the \cii\ luminosity is an effective
tracer of the star formation rate in low redshift starbursts
\citep{De-Looze:2014,Sargsyan:2014}. In addition to the far-IR
luminosity and SED (largely constrained by UV, rather than FIR,
fluxes) this gives three independent measures of the SFR in this $z>6$
galaxy. For the \cii\ luminosity we use the relation SFR
$(M_\odot\,{\rm yr}^{-1}) = 1.0\times10^{-7}L_{\rm [CII]} \,
(L_\odot)$ \citep{Sargsyan:2014}. 

We calculate the far-IR luminosity $L_{\rm FIR}$ (integrated over
  rest-frame 42.5 to 122.5\,$\mu$m) from the observed 1.2\,mm
continuum assuming a greybody spectrum with dust temperature,
$T_{\rm d} =30$\,K and emissivity index, $\beta=1.6$. We use
$T_{\rm d} =30$\,K because that is the dust temperature for similar
FIR-luminosity galaxies at low-redshift \citep{Symeonidis:2013}. We
note that a higher dust temperature of $T_{\rm d} =45$\,K would
increase the $L_{\rm FIR}$ value by a factor of about three. To
convert from $L_{\rm FIR}$ to SFR we use the relation SFR
$(M_\odot\,{\rm yr}^{-1})=1.5\times10^{-10}L_{\rm FIR}\, (L_\odot)$
appropriate for a Chabrier IMF \citep{Carilli:2013}. 

At these redshifts the Cosmic Microwave Background (CMB) has a
temperature of 19\,K and can potentially bias measurements of dust
continuum luminosity. There are two competing effects: (i) a high CMB
background against which the continuum is measured, and (ii) an
increase in the dust temperature due to heating by the CMB. Since we
only have one continuum point and no constraints on dust temperature,
we cannot make an accurate correction for these effects, but note that
according to the analysis of \citet{da-Cunha:2013} the two effects are
of comparable size and opposite sign for likely dust temperatures at
this redshift, so we make no correction.

\begin{figure*}[t]
\includegraphics[angle=0,scale=0.60]{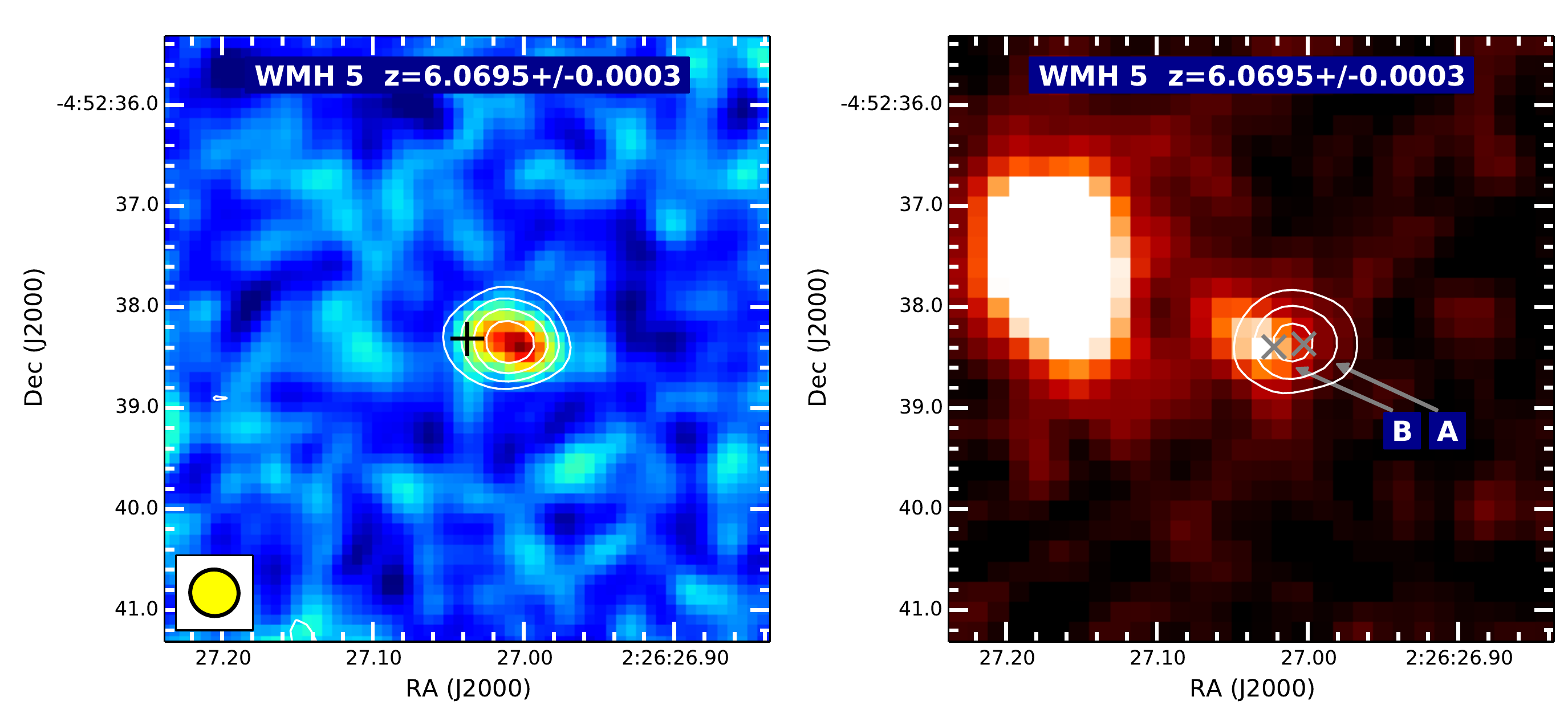}
\caption{{\it Left:} The background image is ALMA integrated \cii\
  line map of WMH 5. White contours show the clear 1.2\,mm dust
  continuum detection (contour levels
  2,\,3,\,4,\,5\,$\sigma$\,beam$^{-1}$). The NIR centroid (black plus
  symbol) is significantly offset to the east.\\ {\it Right:} The
  background is the $zYJ$ NIR image. Contours show the ALMA \cii\
  emission from the left panel at contour levels
  3,\,6,\,9\,$\sigma$\,beam$^{-1}$. Grey crosses labeled `A' and `B'
  correspond to the centroids of the two \cii\ velocity
  components. The rest-frame UV continuum is consistent with `B', the
  component with lower  \cii\ luminosity and velocity width. There is a foreground galaxy further east of the system.} 
\label{fig:mapswmh5}
\end{figure*}

\begin{figure}
\hspace{-0.1cm}
\vspace{0.1cm}
\includegraphics[angle=0,scale=0.32]{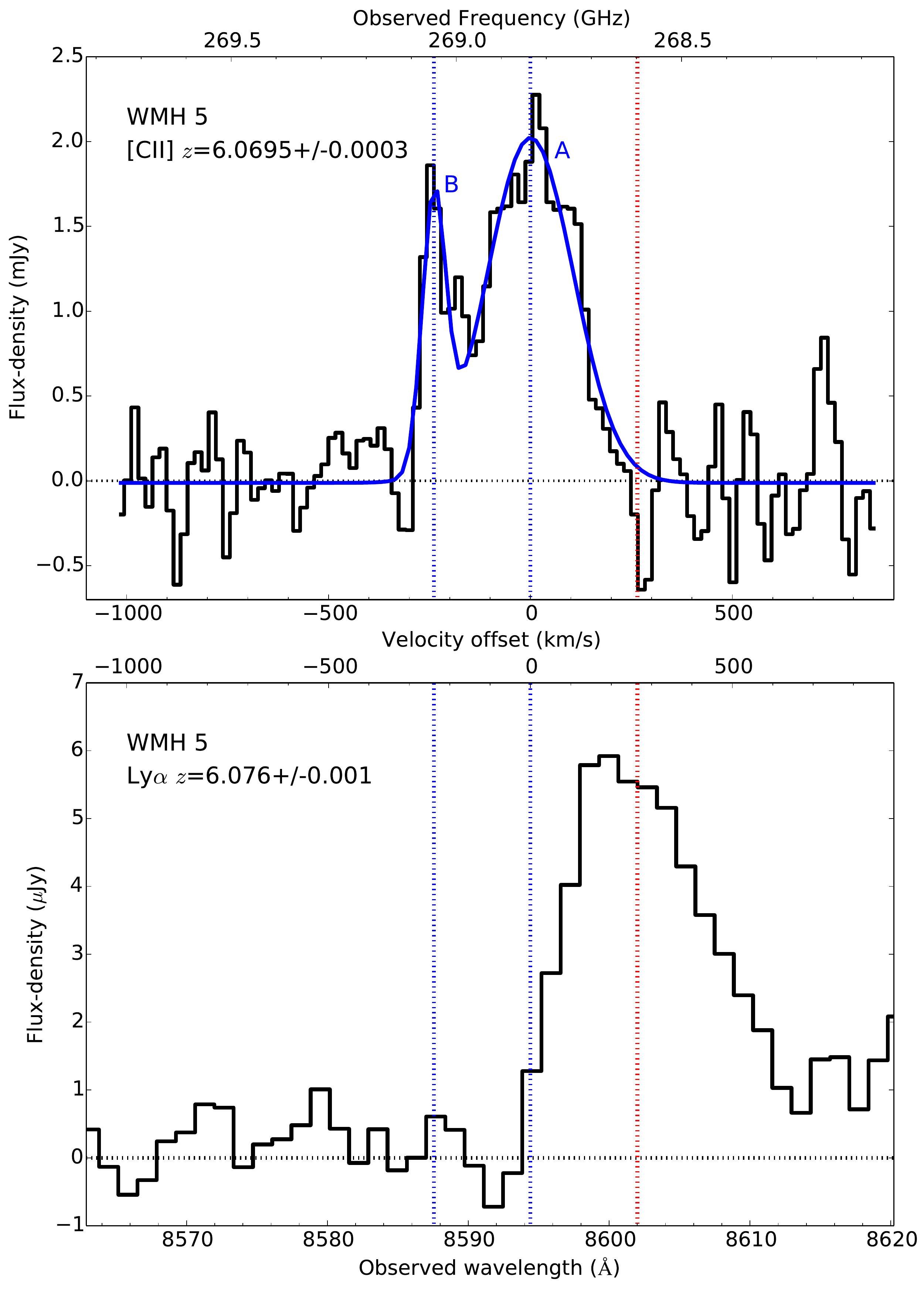}
\caption{ALMA \cii\ (upper) and Gemini GMOS \lya\ (lower) spectra of WMH 5 (see Figure \ref{fig:linespecclm1} for details). The \cii\ data have been continuum-subtracted in the image plane. For this galaxy there are two \cii\  velocity components (`A' and `B'), that are each well fit by a Gaussian. The systemic redshift is taken to be the Gaussian peak of `A' as this is the most luminous and highest width component. The \lya\ spectrum is asymmetric and the flux-weighted centroid offset by $+265 \pm 52$ km\,s$^{-1}$ from `A' and $+504 \pm 52$ km\,s$^{-1}$ from `B'.}
\label{fig:linespecwmh5}
\end{figure}

All three SFR estimates are listed in Table 1. For CLM 1 we find that
SFR$_{\rm SED}$ and SFR$_{\rm [CII]}$ are comparable, but SFR$_{\rm
  FIR}$ is lower than SFR$_{\rm [CII]}$ (see Section \ref{ciifir}) and
lower than SFR$_{\rm SED}$, as expected from the low dust contribution
to the SED in Figure \ref{fig:sedclm1}. Previous studies have shown
that SFR$_{\rm FIR}$ may be unreliable as a tracer of the total SFR in
very low dust and/or metallicity galaxies
\citep{Ouchi:2013,Fisher:2014,Ota:2014}. An alternative is that a
higher dust temperature of $T_{\rm d} =45$\,K would raise SFR$_{\rm
  FIR}$ to a level comparable with SFR$_{\rm SED}$ and SFR$_{\rm
  [CII]}$. Future observations at shorter wavelength are critical to
constrain the full IR SED of high-$z$ galaxies and constrain the dust
temperature.

\subsection{WMH 5}

The ALMA and NIR images for WMH 5 are shown in Figure
\ref{fig:mapswmh5}. There is a much more firm 1.2 mm continuum
detection in this source at significance $6\,\sigma$. The beam is more
circular for WMH 5 than for CLM 1 and both the \cii\ and 1.2\,mm
continuum are clearly spatially extended at PA$\approx 80$. The \cii\
emission centroid shows a significant offset from the NIR emission of
$0\farcs4$. The NIR emission is consistent with a point source. The
\lya\ emission is also spatially unresolved along the slit direction.

\begin{figure}
\hspace{-0.6cm}
\includegraphics[angle=0,scale=0.62]{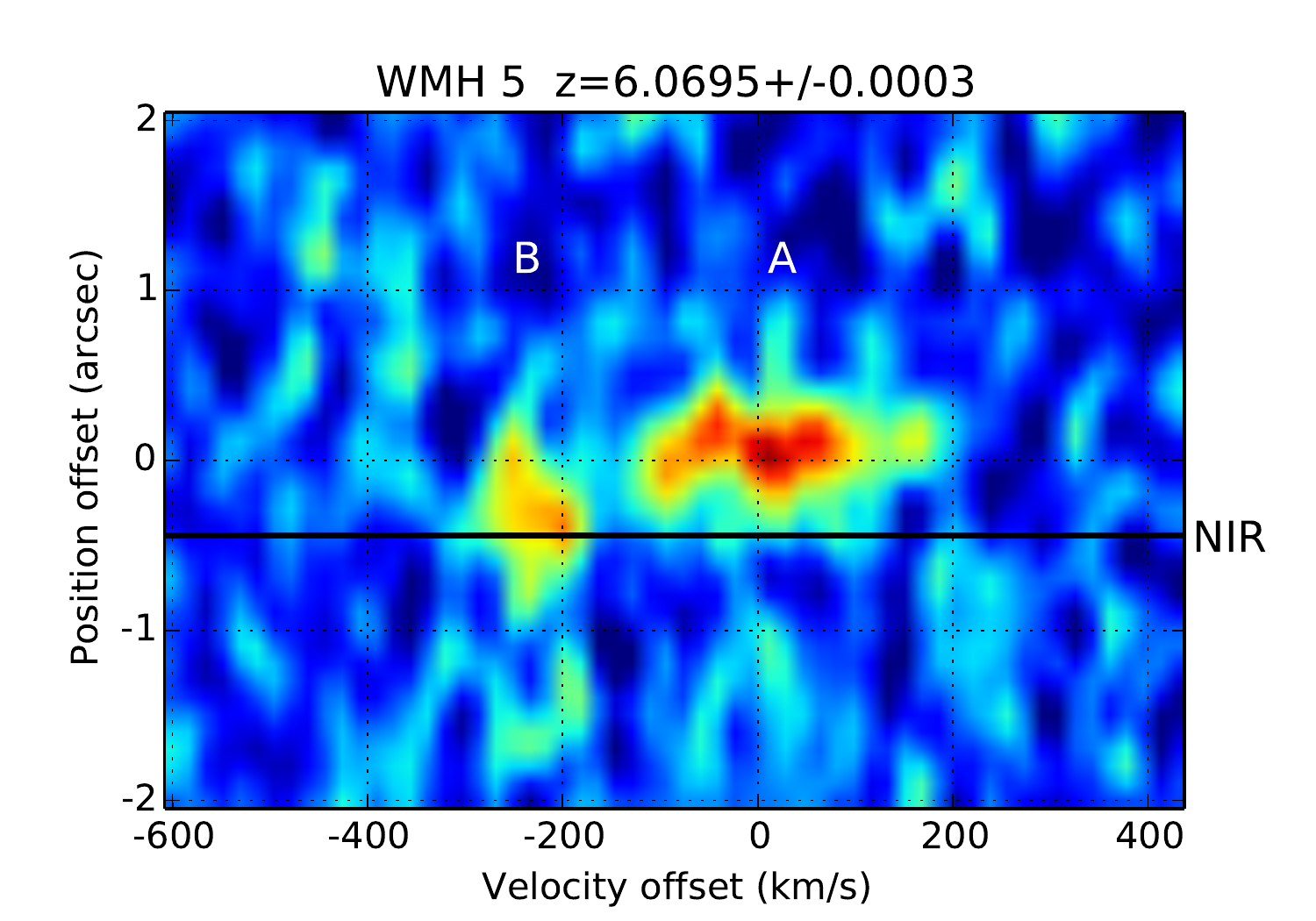}
\caption{Position-velocity map for the \cii\  emission in WMH 5. The NIR galaxy centroid is located approximately along the major axis of the \cii\ emission and its position is identified by the solid black line labelled NIR. The two \cii\ components, `A' and `B' are clearly separated in position and velocity, with `B'  close to the NIR centroid, as seen in Figure \ref{fig:mapswmh5}.} 
\label{fig:pvmapwmh5}
\end{figure}

Figure \ref{fig:linespecwmh5} shows the \cii\ spectrum of WMH 5. The
line can be split into two Gaussian components. We label the component
with higher flux and larger linewidth (FWHM=$251$\,km\,s$^{-1}$) as
`A' and the other component as `B' (FWHM=$68$\,km\,s$^{-1}$). Assuming
the linewidths trace mass gravitationally we identify the systemic
redshift of the main galaxy WMH 5 with the velocity of `A'. To
determine the nature of this multiple velocity component system we
show in Figure \ref{fig:pvmapwmh5} a position--velocity diagram along
the major axis of \cii\ emission. The separation of the two components
in velocity is clear, but we also find a spatial offset
of $0\farcs3$ (2\,kpc) between the peaks of the two components. In Figure
\ref{fig:mapswmh5} we mark the centroids of components `A' and `B'
separately to highlight how those positions compare with each other
and the NIR emission. Both \cii\ components are marginally resolved
with intrinsic major axis sizes of $0\farcs48 \pm 0\farcs1$ and
$0\farcs4 \pm 0\farcs1$ for `A' and `B', respectively. We determine
dynamical masses from the \cii\ sizes and FWHM of $M_{\rm dyn}=
1.7\times10^{10} / \sin^2i \,M_\odot$ for A and $M_{\rm dyn}=
1.0\times 10^9 / \sin^2i \,M_\odot$ for B.

\begin{figure}
\hspace{-0.35cm}
\includegraphics[angle=0,scale=0.48]{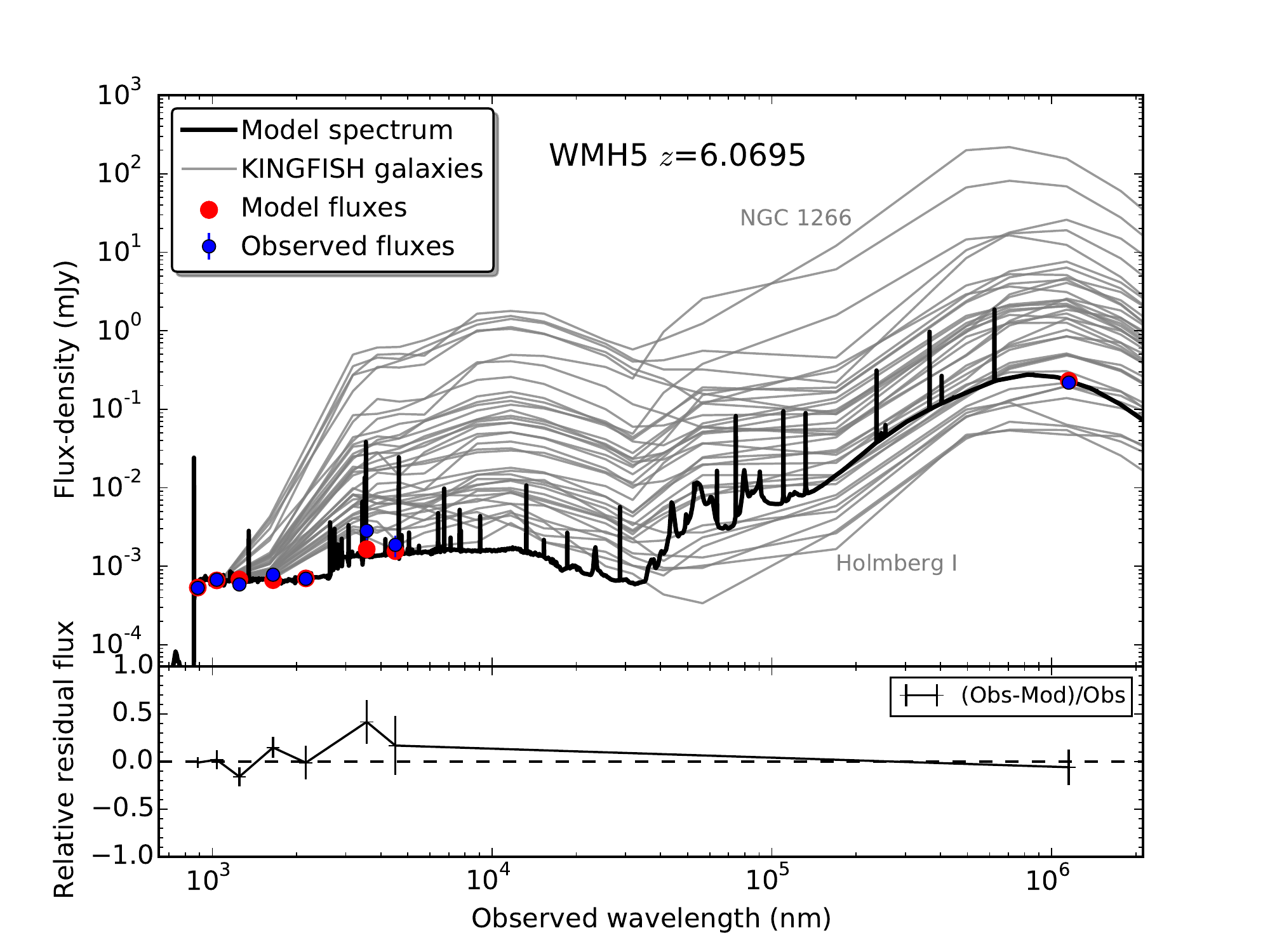}
\caption{Observed-frame optical to far-IR SED of WMH 5 (blue circles), best-fit  {\small CIGALE} model and SEDs of nearby  KINGFISH survey galaxies (see Figure \ref{fig:sedclm1} for details). WMH 5 has a SED most similar to nearby dwarf irregulars.}
\label{fig:sedwmh5}
\end{figure}

We carry out SED-fitting using CIGALE for the entire WMH 5 system. The
observed and modelled SEDs are shown in Figure \ref{fig:sedwmh5}. As
for CLM 1, the photometry is well fit by a single stellar population
whose attenuated UV emission is re-radiated in the FIR. The highest
likelihood parameters are a stellar mass of $2.3 \times 10^{10}
\,M_\odot$, SFR\,$=43 \,M_\odot {\rm yr}^{-1}$ and dust attenuation of
E(B-V)=0.05. The higher dust attenuation in WMH 5 is constrained by
the redder UV spectral slope and consistent with the higher $L_{\rm
  FIR}$. However, in this system we know that there is a spatial
offset between the UV and FIR, so such a simple scenario of
attenuation and re-emission is not physically plausible.

From this analysis the system appears to be an on-going merger of two
galaxies. `B' is apparently lower mass but is spatially coincident
with the NIR emission. This is similar to some lower redshift ULIRG
mergers \citep{Chapman:2004} with an optically obscured component that
dominates the FIR emission and a lower $L_{\rm FIR}$ component less
obscured by dust. Similar cases of most of the \cii\ associated with optically-faint components have been reported at $z>5$ by \citet{Capak:2015,Maiolino:2015}, however in our case component `A' also dominates the FIR continuum and has a large linewidth so is not some peripheral, satellite gas cloud. The stellar mass estimate of $\sim 2 \times
10^{10} \,M_\odot$ is $> 5$ times larger than the dynamical mass of
`B' of $1.0\times 10^9 / \sin^2i \,M_\odot$, unless the inclination
angle $i<30^{\circ}$.  Either `B' is a disk viewed extremely face-on
or it is not responsible for all the stellar mass. 

For WMH 5 the SFR estimates from the far-IR, \cii\ and SED-fitting in
Table 1 are all within a factor of two. This is rather surprising
given the complex nature of the system. The millimeter and UV SFR have
similar values, but their spatial displacement indicates that they are
not probing the same star-forming regions, so the true SFR should be
the sum of both components.

\section{The \cii\ -- far-IR luminosity relation}
\label{ciifir}

\begin{figure}
\hspace{-0.35cm}
\includegraphics[angle=0,scale=0.42]{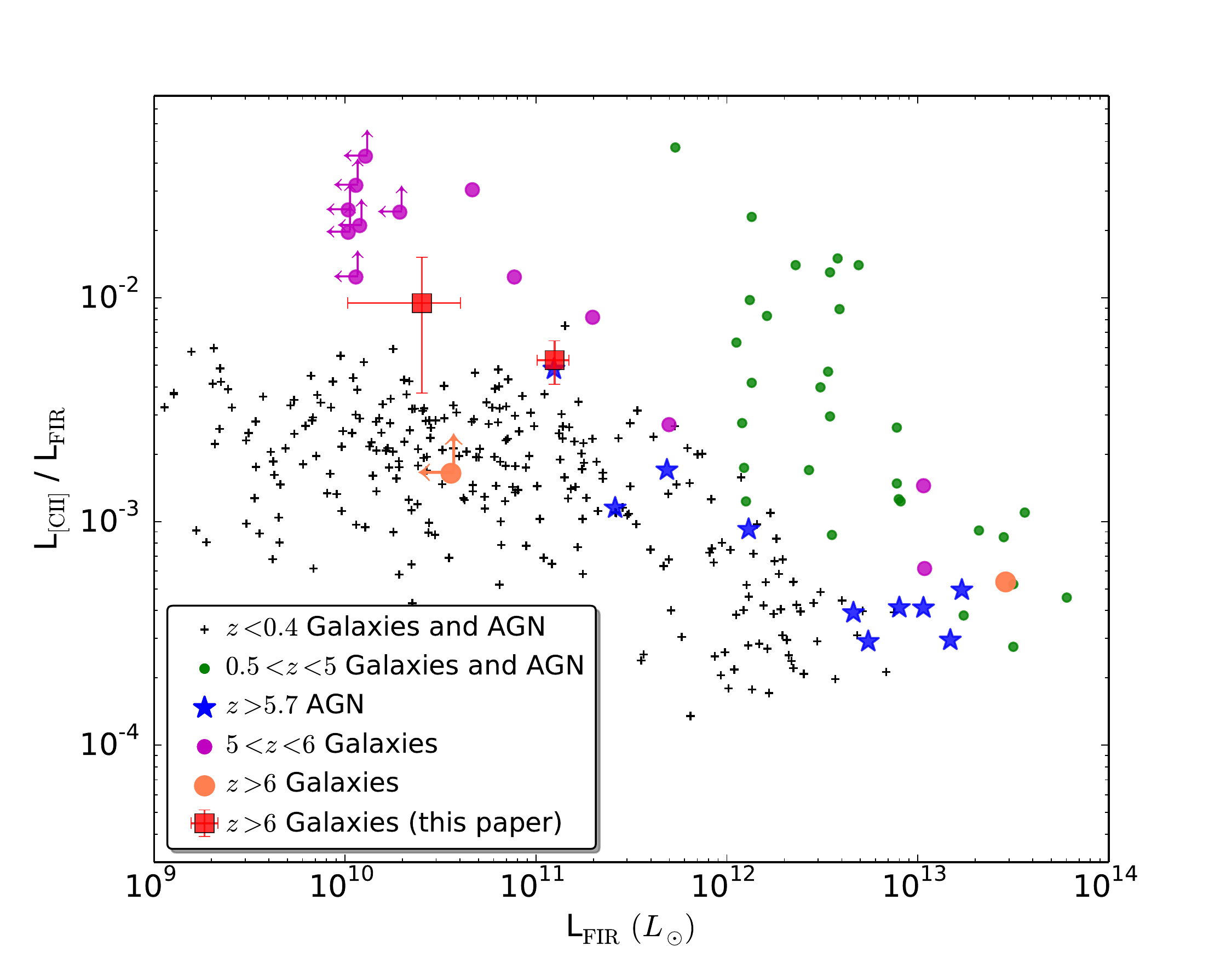}
\caption{Ratio of \cii\ to far-IR (42.5 to 122.5\,$\mu$m) luminosity
  versus far-IR luminosity. The two $z>6$ LBGs from this paper are
  plotted as red squares with error bars. Other galaxies detected in
  the \cii\ transition at $z>6$ plotted as orange circles are
  (left-to-right) BDF3299 Clump A at $z=7.107$ \citep{Maiolino:2015}
  and HFLS 3 at $z=6.337$ \citep{Riechers:2013}. Galaxies at $5<z<6$
  (magenta circles) are from
  \citet{Rawle:2014,Riechers:2014,Capak:2015} where the Capak et
  al. total IR luminosities have been divided by a factor 1.75 to give
  $L_{\rm FIR}$. AGN-hosting galaxies at $z>5.7$ are shown with blue
  stars
  \citep{Maiolino:2005,Venemans:2012,Wang:2013,Willott:2013,Willott:2015}. Galaxies
  (including some AGN) at $0.5<z<5$ (green circles) are from
  \citet{Brisbin:2015} and the compilation of \citet{De-Looze:2014}
  and at $z<0.4$ (black plus symbol) from the compilation of
  \citeauthor{Gracia-Carpio:2011} (2011 and in prep.). The $z>6$ LBGs
  CLM 1 and WMH 5 have ratios comparable to $z>5$ galaxies of similar
  $L_{\rm FIR}$, somewhat higher than at $z<0.4$.}
\label{fig:lciilfir}
\end{figure}

For both \cii\ and FIR luminosities to act as reliable star formation
rate indicators requires that the ratio of the two behave in a
predictable manner without too high a scatter. At low-redshift most
galaxies have ratios in the range $10^{-3}$ to $10^{-2}$ with a fairly
linear relationship between the two luminosities
\citep{De-Looze:2014,Sargsyan:2014}. The exception to this is in the
ULIRG regime where a deficit of \cii\ luminosity is usually observed,
thought to be related to extreme densities and temperatures in
circumnuclear starburst regions
\citep{Farrah:2013,Magdis:2014,Gonzalez-Alfonso:2015}. Figure
\ref{fig:lciilfir} plots the low redshift data with small black
circles.

At higher redshift ($z>4$), the first \cii\ measurements were made in
very high $L_{\rm FIR}$ sources and showed low \cii /FIR ratios comparable
with nearby ULIRGs \citep{Maiolino:2005,Iono:2006}. Subsequent work at
redshifts between 1 and 7 revealed a wide range of ratios from
$10^{-4}$ to $10^{-1}$. At the low end there are high-luminosity $z>5.7$
quasars \citep{Wang:2013} and hyperluminous infrared galaxies (HyLIRGS; $L_{\rm
  FIR}>10^{13} \,L_\odot$) \citep{Riechers:2013,Riechers:2014}. At the
opposite end of the range, high \cii /FIR ratios up to $10^{-1}$ have
been found in some massive $1>z>2$ ULIRGs lying on the {\it
  main-sequence} of star-forming galaxies
\citep{Stacey:2010,Brisbin:2015}. In these galaxies, the
star-formation is more spatially extended than in nearby ULIRGs
leading to higher \cii\ luminosities. 

In Figure \ref{fig:lciilfir} we also include recent ALMA observations
of $L_{\rm FIR}<10^{12} \,L_\odot$ objects comprising $z>6$ AGN
\citep{Willott:2013,Willott:2015} and $z>5$ galaxies
\citep{Capak:2015,Maiolino:2015} plus CLM 1 and WMH 5. Our data are
the first non-limit data at $z>6$ in $L_{\rm FIR}<10^{12} \,L_\odot$
galaxies without AGN. For $z>5$ galaxies the \cii /FIR ratios display a broad
range at somewhat higher values than at low-redshift. Ratios of $\sim
10^{-2}$ suggest extended star-formation with low metallicity and an
intense radiation field.

The similarity of high-redshift galaxies hosting quasars and without
quasars gives us confidence that the star formation properties are
only weakly affected by black hole accretion onto moderate mass black
holes of $M_{\rm BH} \sim 10^{8} \,M_\odot$, compared to the very low
\cii /FIR values found in the more luminous (in both accretion and
FIR) $z>5$ $M_{\rm BH} \sim 10^{9} \,M_\odot$ quasars of
\citet{Wang:2013}. We note that the $L_{\rm FIR}<10^{12} \,L_\odot$ $z>6$ AGN have observed ratios
$f_{\rm 1.2mm}/I _{\rm [CII]} $ similar to CLM 1 and WMH 5, their lower
ratios in this figure being due to the factor of 4 higher $L_{\rm
  FIR}$ calculated using a dust temperature of $T_{d}=47\,K$, compared to
$T_{d}=30\,K$ for the galaxies.

\section{\lya\ emission line velocity shifts and interpretation of rapid evolution in $z\approx 7$ \lya\ galaxies}
\label{lyashifts}

Atomic or molecular gas in star forming regions spread throughout
galaxies is ideal for measuring the systemic redshift of the
galaxy. At high redshift such measurements have been used to determine
the ionized bubble sizes surrounding luminous quasars
\citep{Carilli:2010}. In this Section we use this information to
determine the velocity shifts of the observed \lya\ emission from CLM
1 and WMH 5. \lya\ at $z>6$ is the most important probe of the
fraction of neutral hydrogen in the IGM and the process of cosmic
reionization.

In Figures \ref{fig:linespecclm1} and \ref{fig:linespecwmh5} we
plotted the \cii\ and \lya\ spectra on the same velocity scale after
correcting the datasets to the local standard of rest. For both
galaxies the \lya\ line is asymmetric with a broader red wing than
blue and the line center is shifted to the red from the systemic
redshift defined by the \cii\ observations. This highly asymmetric
line shape is characteristic of \lya\ at high-redshift
\citep{Shimasaku:2006} due to neutral hydrogen absorption of the blue
wing. For CLM 1 the measured offset is $+430 \pm 69$ km\,s$^{-1}$ and
the \lya\ flux drops to zero just before the \cii\ line center. For
WMH 5 there is uncertainty in the systemic redshift due to the two
components. If `A', the component with larger \cii\ linewidth and
therefore likely higher mass, is to be identified as the \lya\ systemic
redshift then the offset is $+265 \pm 52$ km\,s$^{-1}$.  However, the
astrometry suggests that `B' is the origin of
the rest-frame UV and therefore \lya\ emission and in this case the
shift is $+504 \pm 52$ km\,s$^{-1}$. For the remainder of this section
we consider both cases as possible.

The emergence of \lya\ from galaxies is a complicated process that
involves resonant scattering off neutral hydrogen and absorption by
dust. The observed \lya\ profiles depend upon factors such as
geometry, gas covering factor, dust and outflow velocity. Irrespective
of the ionization state of the IGM, galaxies tend to show \lya\
profiles offset to the red due to resonant scattering and selective
absorption within galactic-scale outflows \citep[see review
in][]{Dijkstra:2014}. At $z>6$ where the neutral fraction of the IGM
becomes significant, it is expected that IGM absorption of \lya\
decreases the overall \lya\ flux and shifts the line center further to
the red \citep{Laursen:2011}. The very sharp decrease in \lya\ line
strength of LBGs between $z=6$ and $z=7$ has been interpreted as evidence for
a rapid change in IGM neutral fraction at this epoch, in tension with
reionization models that predict a smoother change in neutral fraction
over this short cosmic time \citep{Pentericci:2011,Schenker:2012,Treu:2013}.

The first measurement of \lya\ velocity shifts at redshifts relevant
to the reionization epoch have recently been made by
\citet{Stark:2014a}. These authors used the UV nebular
\ciii$\lambda1909$ doublet, which is a relatively strong line in
young, low-mass star-forming galaxies \citep{Stark:2014}, to determine
the systemic redshifts of two strong \lya\ emitters. The
gravitationally-lensed $z=6.027$ galaxy A383-5.2 has a high \lya\
equivalent width with strong asymmetry. \citet{Stark:2014a} quote the
\lya\ {\it peak} velocity offset as $120$ km\,s$^{-1}$. For comparison
with our velocity offsets and previous measurements at lower redshift
we determine from their Figure 5 an approximate {\it centroid-based} velocity offset for this galaxy
of $150 \pm 30$ km\,s$^{-1}$. The $z=7.213$ galaxy
GN-108036 has a more moderate \lya\ equivalent width and a much lower
significance detection of \ciii\ was obtained by \citet{Stark:2014a}
with a very low and uncertain velocity offset of $-37 \pm 113$
km\,s$^{-1}$.

\begin{figure}
\vspace{0.06cm}
\hspace{-0.35cm}
\includegraphics[angle=0,scale=0.42]{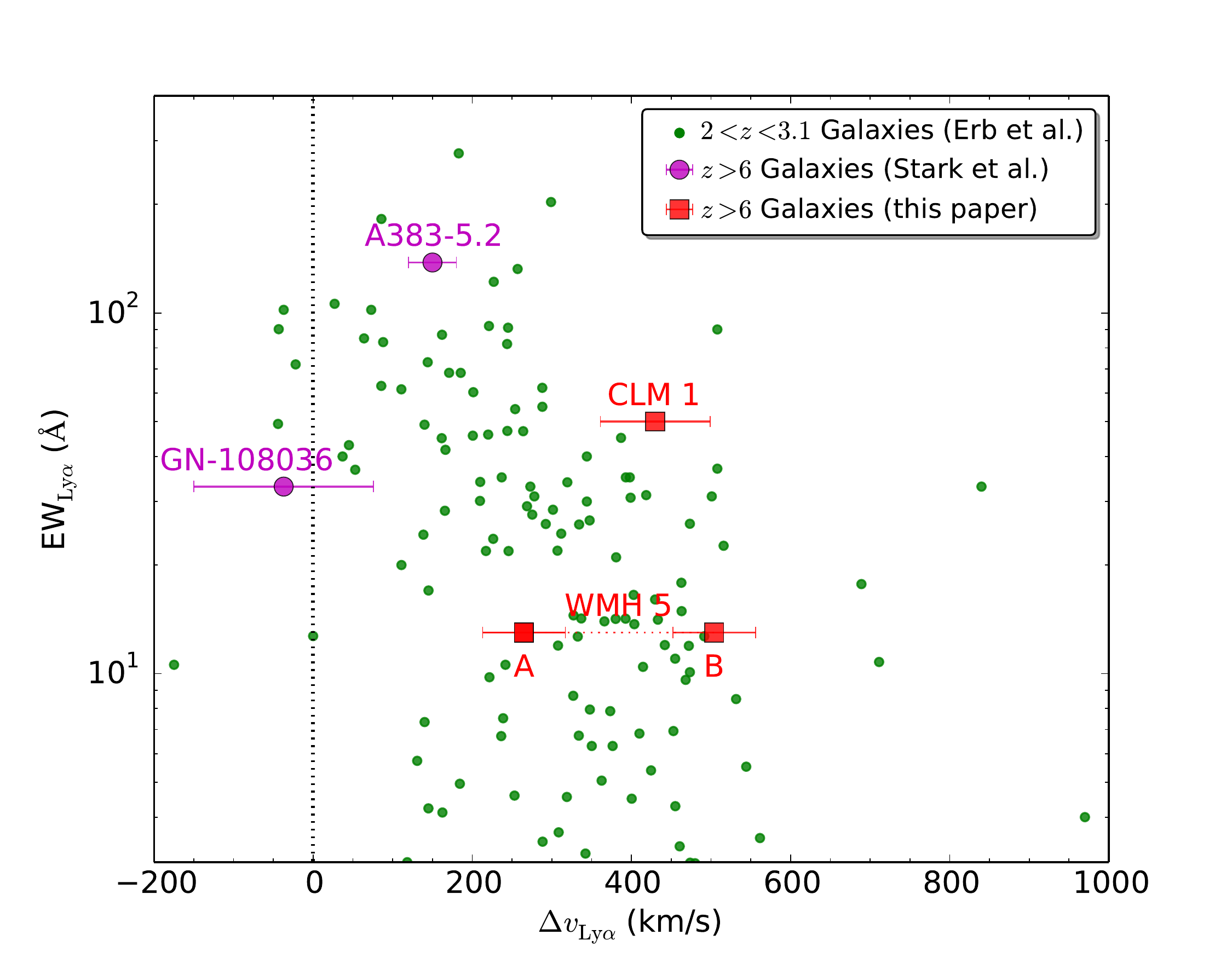}
\caption{Rest-frame equivalent width of the \lya\ emission line versus the velocity offset between \lya\ and the systemic redshift. The two $z>6$ LBGs from this paper are plotted as red squares with error bars. For WMH 5 the two \cii\ components, `A' and `B', provide two estimates of the velocity offset connected by a dotted line. Magenta circles show the only other two galaxies at such high redshifts with measured \lya\ velocity offsets \citep{Stark:2014a} based on the offset to the \ciii$\lambda1909$ UV emission line. A comparison sample of galaxies at intermediate redshift from \citet[also Strom et al. in prep.]{Erb:2014} is plotted as green circles. If the systemic redshift of the WMH 5 \lya\ emission is given by `B', then the two $z>6$ galaxies from this paper lie at the large velocity shift end of the distribution defined by the intermediate redshift sample.} 
\label{fig:lyashifts}
\end{figure}

\citet{Stark:2014a} noted that these offsets are small compared to
those of LBGs at lower redshift and this may be due to different
physical conditions at high redshift. In particular there is known to
be a negative correlation between velocity shift and \lya\ equivalent
width \citep{Shibuya:2014,Erb:2014} and \lya\ strengths increase with
redshift up to $z=6$ \citep{Stark:2011}. Since \lya\ more easily
escapes from galaxies via the wings than the line center, a decrease
in velocity shift at high-redshift could be partially responsible for
the sudden decrease in \lya\ emitters at this
epoch. \citet{Choudhury:2014} showed that a negative evolution of
velocity shift with redshift can match the observed \lya\ equivalent
width distribution evolution to smooth reionization history models
(see also \citeauthor{Bolton:2013} 2013).

In Figure \ref{fig:lyashifts} we compare the \lya\ velocity shifts
measured for CLM 1 and WMH 5 to those of A383-5.2 and GN-108036. We
choose to plot the shifts as a function of observed \lya\ rest-frame
equivalent width because of the correlation between velocity shift and
\lya\ equivalent width. In addition to these $z>6$ galaxies we show
for comparison a sample selected at lower redshift ($2<z<3.1$) as LAEs
and LBGs \citep[Strom et al. in prep.]{Erb:2014}. For the lower
redshift sample, spectroscopic equivalent widths are plotted when
available, otherwise photometric equivalent widths are used. Several
things are evident from looking at Figure \ref{fig:lyashifts}. The
shifts for CLM 1 and WMH 5 are larger than for A383-5.2 and
GN-108036. None of the high-redshift galaxies have shifts at odds with
those observed at $2<z<3.1$ when compared to galaxies with similar
\lya\ equivalent width. If we adopt `B' as the systemic redshift for
WMH 5, then both galaxies in this paper have \lya\ velocity shifts at
the high end of the distribution, consistent with additional IGM
absorption at higher redshift as predicted by the models of
\citet{Laursen:2011}.  Our analysis shows no observational evidence
for velocity shifts that decrease as a function of redshift, at a
given \lya\ equivalent width. Obviously a much larger sample of
observed high-redshift galaxies compared to realistic simulations
incorporating radiative transfer would be required to use the \lya\
velocity shift distribution to constrain the details of reionization.

\section{Conclusions}

We have presented ALMA \cii\ line and dust continuum detections of two
UV-luminous LBGs at redshift $z>6$. These detections were made in
relatively short integrations in {\it Early Science} operations
providing hope that other galaxies in the reionization epoch with
lower SFR will be detectable with the full ALMA array. Our results, in
accord with other recent ALMA observations \citep{Capak:2015}, confirm
that, despite unexpectedly low $L_{\rm FIR}$, the increase in the \cii
/FIR ratio at high-redshift leads to \cii\ lines that are bright and
very useful tracers of the ISM, as predicted by \citet{Walter:2008}.  The
resolution of the merger of WMH 5 with a fairly compact ALMA array 
illustrates the power of \cii\ line observations to understand the
details of star formation and galaxy assembly at this epoch.

CLM 1 and WMH 5 were selected for study because their UV luminosity,
stellar mass and redshift are comparable to the galaxy {\it Himiko}
which was unexpectedly undetected by ALMA \citep{Ouchi:2013,Ota:2014}.  The
main difference between our target galaxies and {\it Himiko} (and
most other high-$z$ non-detections) are our relatively low
\lya\ equivalent widths. Since the \lya\ line strength decreases with
increasing dust (and hence metallicity), this could explain the
difference between these results. We note that the recent success with
ALMA at $5<z<6$ is also for UV-luminous galaxies with relatively weak \lya\
\citep{Capak:2015}.

However, the UV slopes of our galaxies are fairly blue, indicating
little dust absorption of the UV photons and we do observe a deficit
of FIR photons compared to expectations, as shown by the SEDs that are
similar to nearby dwarf irregulars, not with nearby galaxies with
similar SFRs of tens of solar masses per year. Future observations at
shorter wavelength are critical to constrain the full IR SED of
high-$z$ galaxies and constrain the dust temperature. Are they very
cool like low metallicity dwarfs, despite the higher UV photon density
and CMB temperature? Determining this is important as the full IR SED
is required to derive more accurate values of SFR$_{\rm FIR}$ to
determine whether this is a more or less accurate SFR indicator at high-z than
SFR$_{\rm [CII]}$ or SFR$_{\rm UV}$.

We have for the first time used millimeter lines to determine \lya\
line velocity offsets from systemic for galaxies in the reionization
era. We find that the shifts in our two galaxies are comparable to the
high end of the distribution at lower redshift, consistent with
predictions from \lya\ line asymmetry due to neutral
hydrogen absorption \citep{Laursen:2011}. More detailed studies in the
future of spatially-resolved kinematics of \cii\ and \lya\ in
such galaxies will allow stronger constraints to be placed on the
escape of \lya\ emission from galaxies and its subsequent absorption
by the IGM.

%% Included in this acknowledgments section are examples of the
%% AASTeX hypertext markup commands. Use \url without the optional [HREF]
%% argument when you want to print the url directly in the text. Otherwise,
%% use either \url or \anchor, with the HREF as the first argument and the
%% text to be printed in the second.

\acknowledgments

Thanks to staff at the North America ALMA Regional Center for processing the ALMA data, Jean-Gabriel Cuby for providing the optical spectrum of CLM 1, Dawn Erb for providing unpublished data on intermediate redshift Lyman Break Galaxies, and Peter Capak and Caitlin Casey for interesting discussions. This paper makes use of the following ALMA data: ADS/JAO.ALMA\#2013.1.00815.S. ALMA is a partnership of ESO (representing its member states), NSF (USA) and NINS (Japan), together with NRC (Canada) and NSC and ASIAA (Taiwan), in cooperation with the Republic of Chile. The Joint ALMA Observatory is operated by ESO, AUI/NRAO and NAOJ. The National Radio Astronomy Observatory is a facility of the National Science Foundation operated under cooperative agreement by Associated Universities, Inc.


%% in parentheses, after the keyword, but they will not be verified.

{\it Facility:} \facility{ALMA}.



\bibliography{willott}

\end{document}